\documentclass[a4paper,12pt]{article}
\usepackage[a4paper,width=160mm,top=35 mm,bottom=25mm]{geometry}
\usepackage[dvips]{graphicx}
\usepackage{color}
\usepackage{cite}
\usepackage{hyperref}
\usepackage[cmex10]{amsmath}
\usepackage{amsthm}
\usepackage{amssymb}
\usepackage{algorithmic}
\usepackage{url}
\usepackage{algorithm}
\usepackage{subfigure}
{}
{}
{}
\usepackage{subfigure}
\usepackage{float}
\usepackage{mwe}
\usepackage[detect-all]{siunitx}
\usepackage{graphicx}
\usepackage{capt-of}
\usepackage{booktabs}
\usepackage{varwidth}
\usepackage{amsmath}
\date{}
\begin{document}
\title{Robustness of a Biomolecular Oscillator to Pulse Perturbations}
\author{Soumyadip Banerjee, Shaunak Sen*\\
  Department of Electrical Engineering\\
  Indian Institute of Technology Delhi\\
  Hauz Khas, New Delhi, India--110016, India\\
  *Email: shaunak.sen@ee.iitd.ac.in       
}
\vspace{-2mm}
\maketitle
\begin{center}
\large{\textcolor{blue}{The paper is a postprint of a paper submitted to and accepted in IET Systems Biology and is subjected to Institution of Engineering and
Technology Copyright. The copy of record is available at IET Digital Library}}\end{center}
\abstract{\noindent Biomolecular oscillators can function robustly in the presence of environmental perturbations, which can either be static or dynamic. While the effect of different circuit parameters and mechanisms on the robustness to steady perturbations has been investigated, the scenario for dynamic perturbations is relatively unclear. To address this we use a benchmark three protein oscillator design - the repressilator - and investigate its robustness to pulse perturbations, computationally as well as using analytical tools of Floquet theory. We find that the metric provided by direct computations of the time it takes for the oscillator to settle after a pulse perturbations is applied, correlates well with the metric provided by Floquet theory. We investigate the parametric dependence of the Floquet metric, finding that the parameters that increase the effective delay enhance robustness to pulse perturbation. We find that the structural changes such as increasing the number of proteins in a ring oscillator as well as adding positive feedback, both of which increase effective delay, facilitates such robustness.These results highlight such design principles, especially the role of delay, for designing an oscillator that is robust to pulse perturbation.}

%
     
\section{Introduction}
Biomolecular circuits underlying cellular behaviour exhibit both steady output, such as a constant concentration of protein in response to an input~\cite{barkai1997robustness}, or dynamic output such as oscillations in protein concentrations, as in circadian rhythms~\cite{peschel2011setting}.
Further, these circuits may be subject to  environmental perturbations, both steady and dynamic, whose effect on the circuit output needs to be attenuated~\cite{stelling2004robustness,shinar2007input}.
Robustness to such perturbations, both steady and dynamic, may be an important objective in the functioning and design of biomolecular circuits.

The robustness of biomolecular oscillations to various perturbations has
been a topic of investigation.
A recent example is a more robust design iteration \cite{potvin2016synchronous} of a delayed negative feedback-based ring
oscillator - the repressilator~\cite{elowitz2000synthetic}, which is a key demonstration of biomolecular circuit design. This was partly achieved by tuning circuit parameters to remove coupled
degradation. Another experimental oscillator design, based on interlinked positive and negative feedback loops, was shown to be robust to a range of conditions such as temperature and inducer levels~\cite{stricker2008fast}. In addition to these experimental studies, there have been computational and analytical work on oscillator robustness. A recent example is a computational search among two-node and three-node oscillators for robust oscillations and their quantification~\cite{woods2016statistical}. Other work has focussed on characterizing the extent of persistence of oscillations for different parameters ~\cite{apri2010efficient,rougemont2007dynamical}. There are also investigations on robustness of oscillator to perturbations which are dynamic in nature. An important source of dynamic perturbation can be molecular noise, which may influence and modify the dynamical properties of an oscillator inside a cell~\cite{zhang2010architecture,gonze2002biochemical,toner2013molecular,mckane2007amplified,qian2002concentration}. Also, we have found preliminary computational evidence of how positive feedback can enhance robustness of the repressilator to pulse perturbation~\cite{banerjee2018attenuation}. Study of these dynamic perturbations are important, as in some cases these are found to reset or even stop an oscillator~\cite{forger2004starting}. These represent important work in understanding biomolecular oscillator designs in cellular contexts, where they may encounter perturbations.

\begin{figure*}[t]
	\centering
	
		\includegraphics[scale=.28]{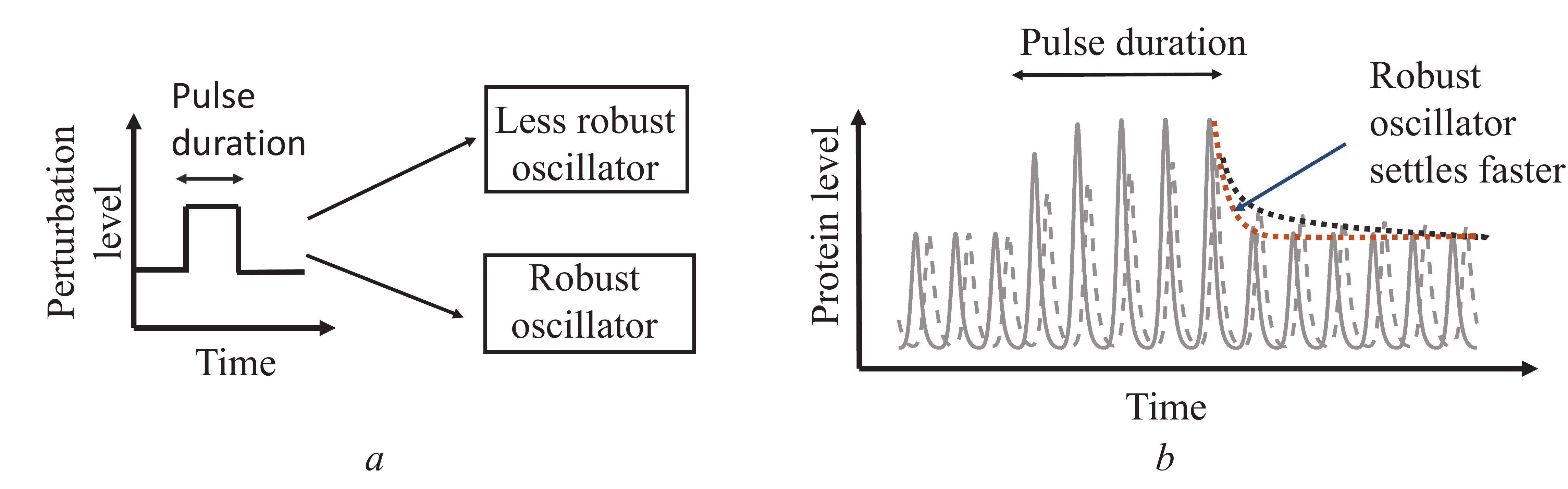}
		\label{fig:concept}

	\caption{Illustration of an oscillator's robustness to pulse perturbation.\newline
 \fontsize{7}{7}\selectfont{\textit{(a)} Schematic showing pulse perturbation applied to two oscillators, one robust and other less robust. \textit{(b)} Schematic showing pulse perturbation applied to two oscillators, one robust and other less robust.}
}
\end{figure*}

There are three striking features in relation to the robustness of oscillations to various perturbations.
One, is how the structure of the circuit influences the robustness. Two, is how, within the same structure, different parameters may have different robustness properties. Three, how the amplitude of the perturbation may also affect the oscillation. Given these, the robustness of oscillator designs to dynamic perturbations such as a pulse is relatively unclear.

Here we ask how to design an oscillator that rapidly attenuates the effect of a pulse disturbance (Figure 1). To address this, we use the repressilator, a benchmark biomolecular oscillator, and the mathematical concept of Floquet multiplier of an oscillating trajectory.  We find that after pulse perturbation, the settling time estimated from the largest Floquet multiplier of an oscillating trajectory strongly correlates with the settling time obtained from a full nonlinear simulation, for both small and large pulse amplitudes. We characterize the dependence of the Floquet multiplier on intrinsic circuit parameters and find that the parameters that increase effective delay can enhance the robustness of oscillation to pulse perturbation. We also considered the effect of structural changes such as the number of nodes in the ring and additional feedback loop, and find that increasing the node number and adding positive feedback, both of which increases effective delay can increase robustness to pulse perturbations.

\section{Results}\label{sec1}

\subsection{Floquet multiplier is a good measure of settling time}\label{sec2.1}
We start with a standard model of the repressilator (Figure 2\textit{(a)},~\cite{elowitz2000synthetic}), a benchmark biomolecular oscillator, and compute the effect of a pulse disturbance on it.  

The standard model is,
\begin{equation}
\begin{aligned}\dot{m}_{P_{1}} & =F(P_{3})-\delta\,m_{P_{1}},\\
\dot{P}_{1} & =k\,m_{P_{1}}-\gamma\,P_{1},\\
\dot{m}_{P_{2}} & =F(P_{1})-\delta\,m_{P_{2}},\\
\dot{P}_{2} & =k\,m_{P_{2}}-\gamma\,P_{2},\\
\dot{m}_{P_{3}} & =F(P_{2})-\delta\,m_{P_{3}},\\
\dot{P}_{3} & =k\,m_{P_{3}}-\gamma\,P_{3},
\end{aligned}
\end{equation}

\noindent where, $m_{P_{1}}$, $m_{P_{2}}$ and $m_{P_{3}}$ are the  messenger RNA (mRNA) concentrations and $P_{1}$, $P_{2}$ and $P_{3}$ are the protein concentrations. $\delta$ is the mRNA degradation rate, $k$ is the average translation rate, $\gamma$ is the protein degradation rate. The repression of all the three protein is modelled by a Hill function $F(P)=\alpha_{0}+\dfrac{\alpha}{1+\left(\frac{P}{K}\right)^{n}} $. Here $\alpha_{0}$ is the leaky expression rate and $\alpha_{0}+\alpha$ is the maximum expression rate, $n$ is the Hill coefficient and K is the dissociation constant. We simulated the above equations numerically in MATLAB using solver ode23s. For a nominal parameter set , the repressilator oscillates with a period of 142 minutes (Figure 2\textit{(a)}).
\begin{figure*}[t!]
	\centering
	\subfigure
	{%
		\includegraphics[scale=.35]{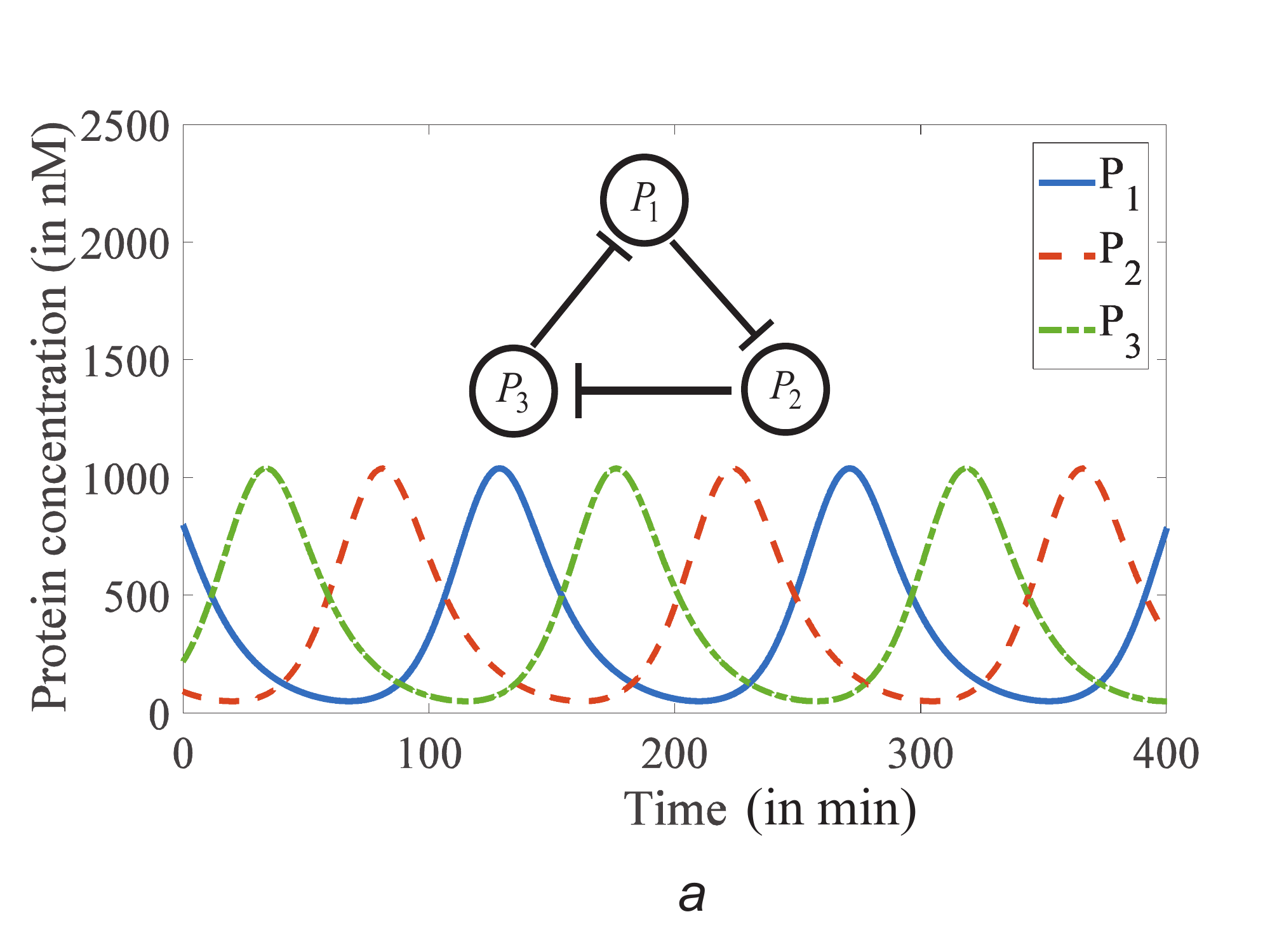}

	}~
\subfigure
	{%
		\includegraphics[scale=.35]{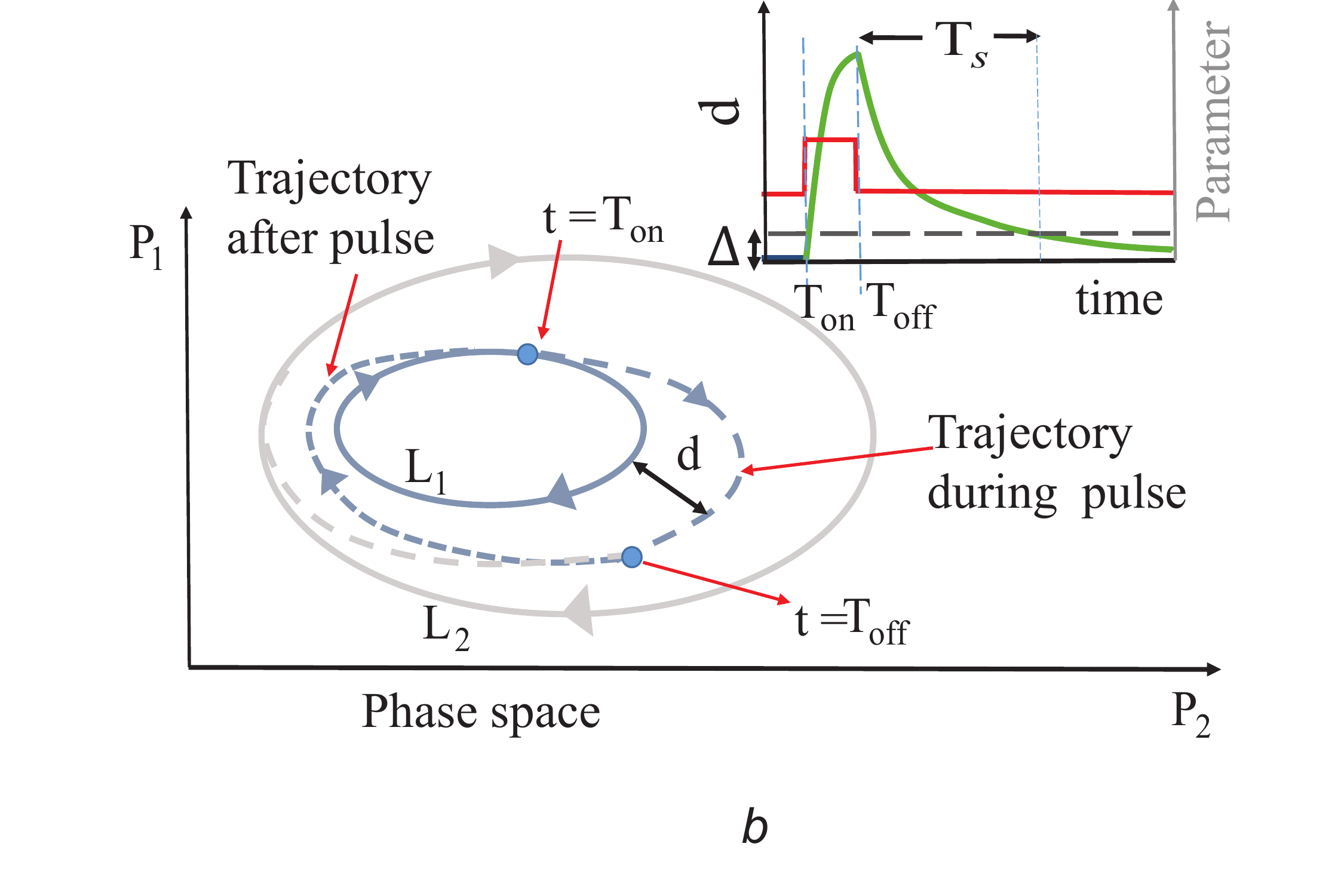}
		
			}	
	\subfigure
	{%
		\includegraphics[scale=.45]{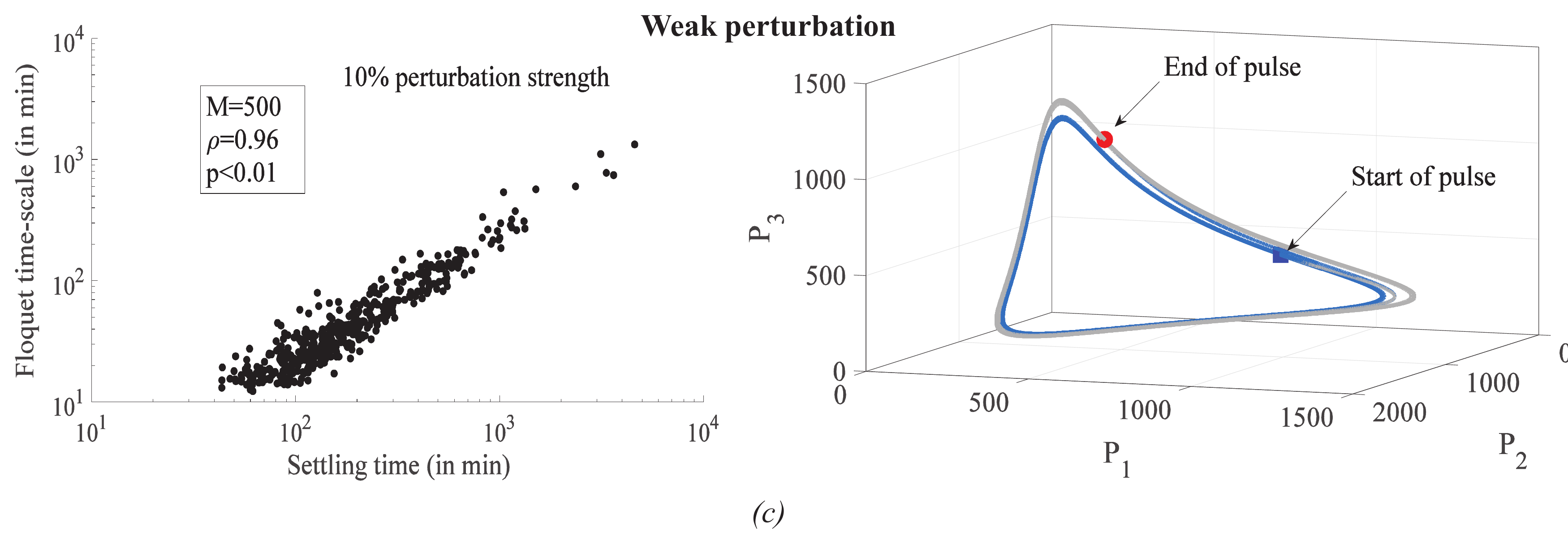}
		
			}	
			\subfigure
	{%
		\includegraphics[scale=.45]{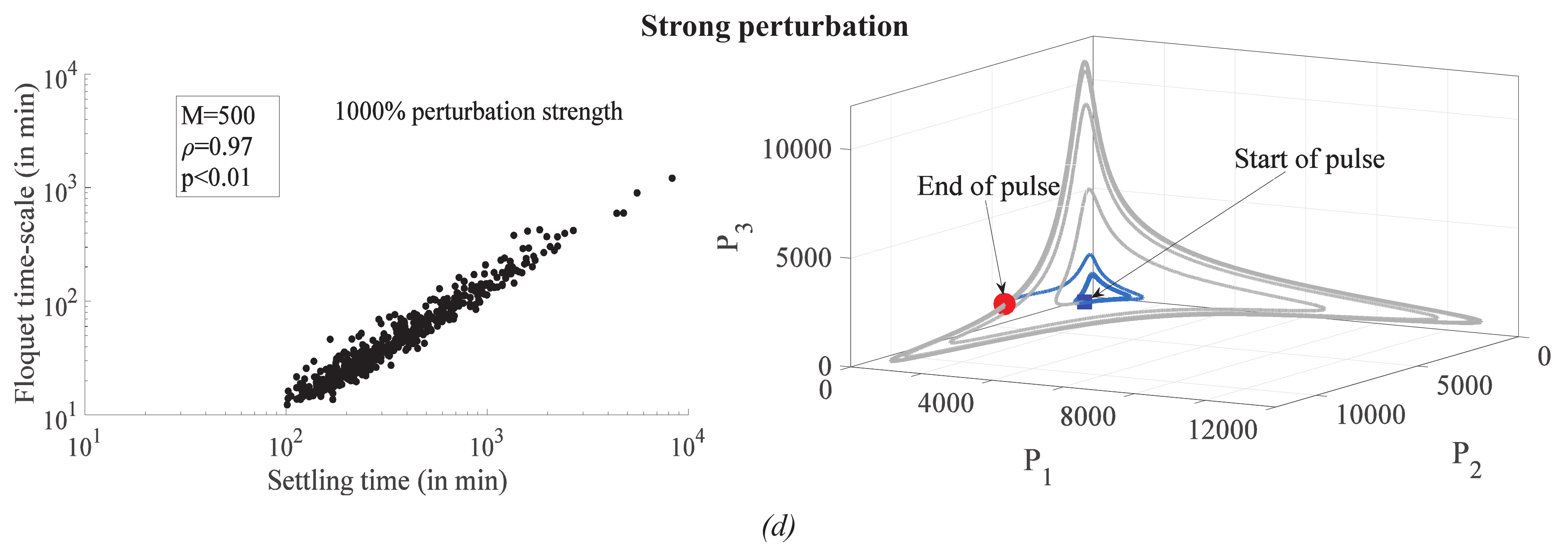}
		
			}

\caption{Dominant Floquet multiplier as a measure of robustness to pulse perturbation.\newline
 \fontsize{7}{7}\selectfont{\textit{(a)} Deterministic simulation of the repressilator model showing the time trace of the three protein concentrations $P_{1}$, $P_{2}$ and $P_{3}$. The simulation is performed using the nominal parameters: promoter strength, $\alpha$ = 10 nM/min (under full induced condition) and $\alpha_{0}$ = $5\times10^{-4}$ nM/min (under repressed condition), $K$ = 50 nM, $n$ = 2 and mRNA half-life = $\frac{1}{\delta}$ = 3 min and protein half-life = $\frac{1}{\gamma}$ = 20 min and $k=5\,\mathrm{mi}\mathrm{n}^{-1}$. \textit{(b)} Approach used to compute the settling time is shown. The solid blue closed curve represents an oscillating trajectory $\mathrm{L_{1}}$ and the grey closed curve with arrows represents the oscillating trajectory $\mathrm{L_{2}}$. The dashed blue line represents the trajectory of the perturbed oscillator undergoing transition between $\mathrm{L_{1}}$ and $\mathrm{L_{2}}$. In the inset the green solid line shows the variation of the minimum distance $\mathrm{d}$ between the perturbed trajectory and $\mathrm{L_{1}}$ with time. The variation of the parameter with time is shown in red. The bound $\varDelta$ is shown as a black dashed line. \textit{(c)} The left figure shows the scatter plot showing the correlation between the Floquet time-scale ($\tau$) and settling time ($\mathrm{T_{s}}$) computed directly for low perturbation strength. Each point represents a periodic orbit corresponding to a parameter set chosen within a range of the nominal parameter set- $\alpha$ =\,\numrange[range-phrase = --]{2}{20}, $K$ =\,\numrange[range-phrase = --]{1}{100}, $n$\,=\,\numrange[range-phrase = --]{2}{4}, $\delta$\,=\,\numrange[range-phrase = --]{.2}{.5}, $k$\,=\,\numrange[range-phrase = --]{1}{10}, $\gamma$\,=\,\numrange[range-phrase = --]{0.01}{.1}. In the inset, M denotes the total number of parameter sets randomly chosen within the above range, $\rho$ denotes the correlation coefficient and p denotes the p-index. For the settling time computation bound, $\mathrm{\varDelta}$ is chosen to be 0.5. The right figure shows the response of repressilator operating at the nominal parameter to weak perturbation. The grey solid line represents the perturbed trajectory of the oscillator when the pulse is on and the blue solid line represent the perturbed trajectory in the post pulse period. The blue  square represent the onset of pulse whereas the red filled circle represent the end of pulse. \textit{(d)} Similar plots as in $\it{c}$, corresponds to high perturbation strength}}

\end{figure*}
A pulse perturbation can be applied by changing a particular parameter for a certain period of time. From a biological point of view, such pulse-type perturbation may arise due to a change in external stimulus such as light~\cite{piechura2017natural} and temperature~\cite{ruoff2001goodwin}, as well as from the output of a pulse-generating biomolecular circuit, such as feedforward loop~\cite{alon2006introduction}, which is an input to the oscillator circuit. The dynamics of the oscillator circuit before and after the application of such a pulse is as follows. Before the pulse, the circuit displays an oscillating trajectory with a period, $T$. This is represented as a periodic waveform in time (Figure 2\textit{(a)}) or a closed curve (Figure 2\textit{(b)}, denoted as $\mathrm{L_{1}}$) in phase space. When the pulse is applied, the circuit moves towards a different oscillating trajectory (Figure 2\textit{(b)}, denoted as $\mathrm{L_{2}}$). When the pulse turns off, the circuit reverts back to the original oscillating trajectory ($\mathrm{L_{1}}$). The settling time required by the perturbed oscillator to converge to this oscillating trajectory can be estimated numerically. We obtain it by computing the time taken for the distance, `d' between the perturbed and the oscillating trajectory to reach a pre-specified value `$\varDelta$'(Figure 2\textit{(b)}) and is denoted as $\mathrm{T_{s}}$. The design objective may be to ensure a small value of $\mathrm{T_{s}}$ so that an oscillator is robust to a pulse perturbation.

The Floquet theory is a natural analytical method to estimate the settling time in such contexts~\cite{strogatz2018nonlinear}. Essentially the Floquet multipliers obtained from it can provide us with the relaxation time constants when an infinitesimally small transient perturbation is made to an oscillating trajectory. We estimate the settling time or the Floquet timescale ($\tau$) for the relaxation process using its relation with the dominant Floquet multiplier given as, (please see, Sec 4.1)

 \begin{equation}\label{floq}
\tau=\frac{T}{ln(1/|\lambda_{max}|)}
\end{equation}

\noindent The Floquet theory has been used in the context of robustness of oscillations to perturbations due to biomolecular noise~\cite{gonze2002biochemical}. Whether it provides a useful metric of settling time for large amplitudes is generally unclear. To address this, we correlated the settling time from the Floquet method with the settling time obtained from direct numerical simulation both for weak and strong amplitude perturbation. The settling time was computed from the dominant Floquet multiplier ($\lambda_{max}$) using standard methods~(\cite{strogatz2018nonlinear,klausmeier2008floquet}, please see Methods Sec 4.2). For the direct computation, the pulse was generated by changing the value of the parameter $\alpha$ to a new level (10\% of its nominal value for small perturbation strength and 1000\% of its nominal value for large perturbation strength) for 10 time periods. To account for the case that a pulse can act at different phases of the cycle, we have given the pulse at a random phase for each parameter set chosen. We find a significant correlation between the settling time obtained directly and through Floquet method for small and large scale perturbations (Figure 2\textit{(c-d)}). This shows that the dominant Floquet multiplier is a good measure of the settling time. Further in Figure 2\textit{(c-d)} we find that the settling time computed directly for the strong perturbation is greater than that for weak perturbation case. A possible explanation for the observed data is that for a large perturbation, the oscillator takes a longer trajectory to converge back to its oscillation as shown in Figure 2\textit{(c-d)} (right).

\subsection{Tuning intrinsic parameters to improve settling time}

Next, we sought to understand the dependence of settling time on the various parameters intrinsic to the repressilator. For this, we plotted the dominant Floquet multiplier and the estimated settling time, as shown in Figure 3. A small value of the dominant Floquet multiplier correlates with a faster settling time indicating the oscillator is robust to pulse perturbation. The period of oscillation also changes as the parameters change (shown in Figure 3).

We note that that the Floquet multiplier either increases or decreases when the parameter $\alpha$, $n$, $\delta$, $k$ and $\gamma$ increase in the selected parameter range. The dependence on $n$ is particularly strong. In the case of $K$, as we increase it, we observe first a decrease and then an increase in the Floquet multiplier value.

The variation of the dominant Floquet multiplier with respect to $\gamma$ and $\delta$ suggests a role for the delay in designing robustness to pulse perturbation. The dominant Floquet multiplier decreases as $\delta$ decreases or as $\gamma$ increases. These parameter regimes correspond to a situation where the transcriptional and translational timescales are of a similar magnitude. If either the transcriptional timescale is faster than the translational timescale (larger $\delta$ relative to $\gamma$) or the translational timescale is slower than the transcriptional timescale (small $\gamma$ relative to $\delta$), then the intermediate states in the ring is lesser~\cite{novak2008design}. This inhomogeneity in the timescale of the intermediate stages lowers the effective delay in the oscillator circuit~\cite{blanchini2018homogeneous} and thus increasing the dominant Floquet multiplier~\cite{ugander2008delay}. This is correlated with a larger Floquet multiplier or a slower settling time. Additional evidence for the role of delay in ensuring robustness to pulse perturbation is obtained from the dependence of the Floquet multiplier on the Hill coefficient $n$. Physically, $n$ represents the cooperativity in the binding events~\cite{whitty2008cooperativity} and as the number of binding events increase, $n$ increases. This means that the effective delay becomes larger as $n$ increases, further correlating with a faster settling time. We note that the same intrinsic parameter has been reported to increase robustness of a biomolecular oscillator to molecular fluctuation \cite{gonze2002biochemical}.

\begin{figure*}[t]
	\centering
	{%
		\includegraphics[scale=.35]{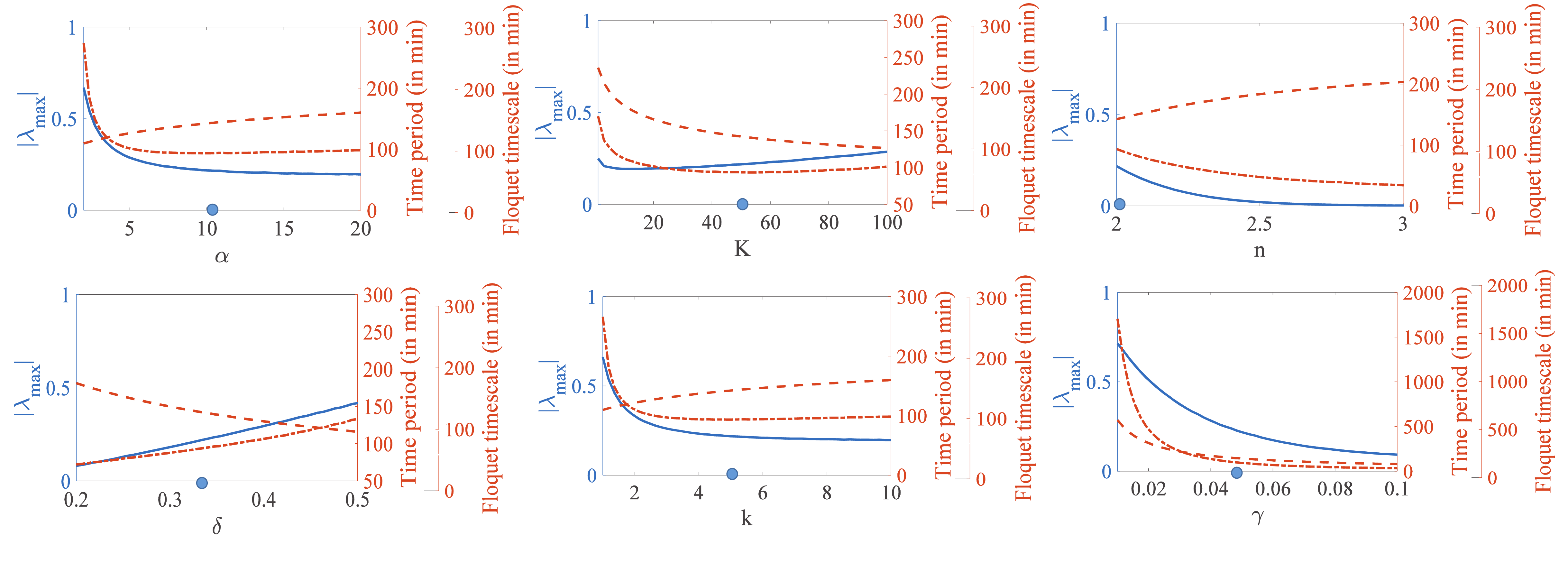}}
		\label{fig:concept}
	\caption{Robustness to pulse perturbation depends on the choice of intrinsic parameters.\newline
 \fontsize{7}{7}\selectfont{In each plot the solid blue line represents the variation of the dominant Floquet multiplier $\lambda_{max}$ with the parameter indicated on the x-axis. The red dot-dashed line represent the settling time $\tau$ predicted by Floquet analysis. The red dashed line represent the time period $T$. The range of parameters are taken around the nominal value which is indicated by a blue filled circle on the x-axis.}
 }
\end{figure*}

\subsection{Structural modifications to improve settling time}

Next, we investigated structural modifications to the repressilator circuit that improve the settling time. Based on the above computations, increasing effective delay may improve settling time. Therefore, we considered the effect of two structural modifications that may increase the effective delay.

\subsubsection{Increasing number of stages in the oscillator ring circuit}\label{ring length}

One way to increase the effective delay is to increase the number of proteins in the repressilator circuit from three to a higher odd number (Figure 4\textit{(a)}). Odd number of nodes in a ring are known to generate oscillations~\cite{niederholtmeyer2015rapid}. The mathematical model for these oscillator is
\begin{equation} 
\begin{aligned}
\dot{m}_{P_{i}} & =F(P_{i-1})-\delta\,m_{P_{i}},\\
\dot{P}_{i} & =k\;m_{P_{i}}-\gamma\,P_{i},
\end{aligned}
\end{equation}
\noindent where, $i\in[1 \; 2 ....N]$, and $F$ is as defined previously. Here $P_{0}$ and $P_{N}$ are identical stages.

 The above model for $N=5$ is compared with the original repressilator model by exploring their parameter spaces randomly and comparing the distribution of Floquet multipliers. The circuit having a lower value of Floquet multiplier is more robust to pulse perturbation. To generate the random sampling sets, we used Latin hypercube sampling ($\it{lhsdesign}$ in MATLAB) and selected 200 parameter sets around the nominal parameter values of the original model (Sec \ref{sec2.1}). On plotting the histogram (Figure 4\textit{(b)}) of the Floquet multipliers, we find that for most of the parameter sets, the 5 node circuit oscillations have a smaller Floquet multiplier compared to that of the 3 node circuit.  The relative difference in the magnitude of the multiplier values can be seen more clearly in the inset presented in Figure 4\textit{(b)}. A typical value of the Floquet multiplier of the 5 node circuit oscillation obtained at the nominal parameter values is 1.18e-04, which is a sharp decrease from that obtained in $N=3$ case ($|\lambda_{max}|=0.22$). In fact, the computation of Floquet multiplier for $N=5$ and beyond may be limited by the numerical resolution of the computational method (please see Sec 4.2). The above result shows that increasing ring size can lead to robust oscillations.      
\begin{figure*}[t!]
	\centering
	\subfigure
	{%
		\includegraphics[scale=.27]{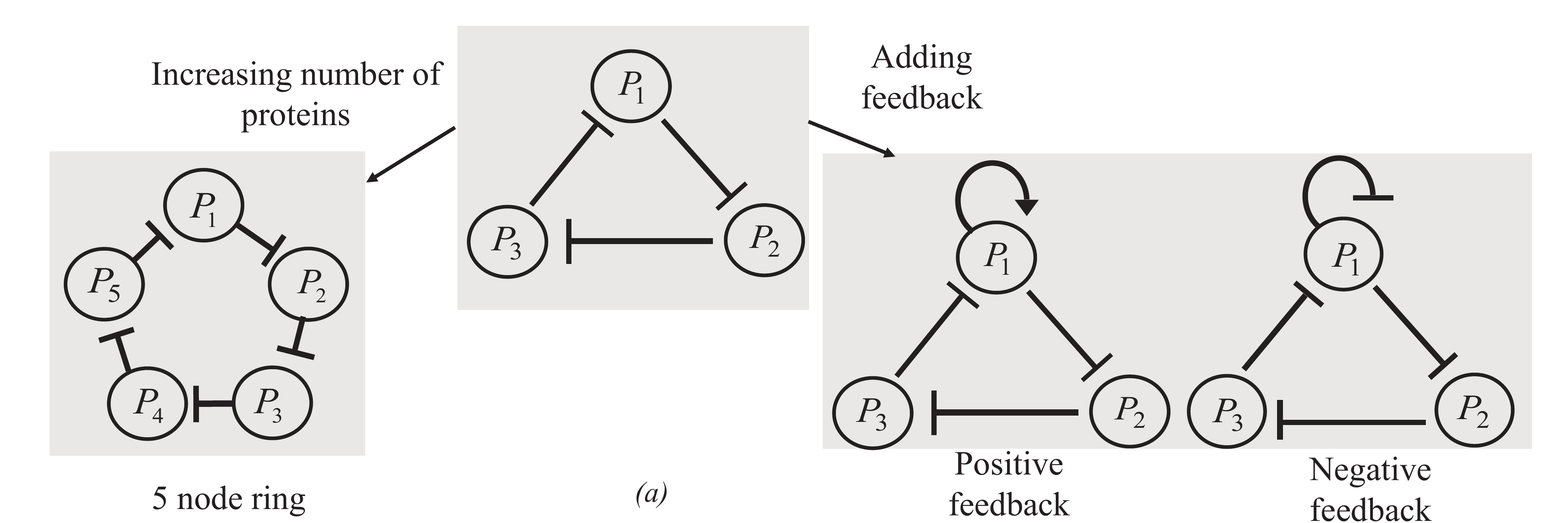}
		\label{feedback_concept}
	}
	
	\subfigure
	{%
		\includegraphics[scale=.32]{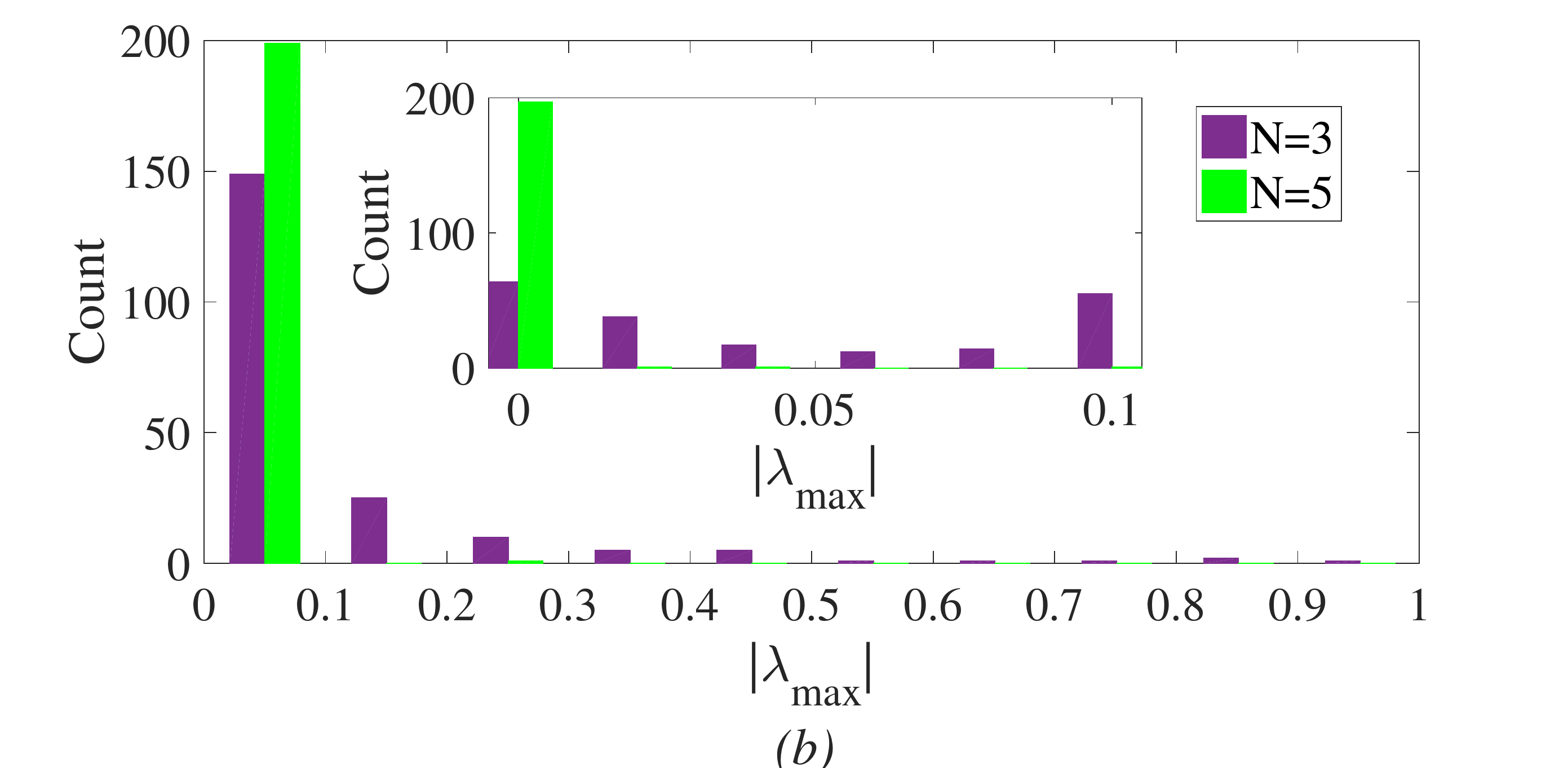}
		\label{strength50}
		}
		~
	\subfigure
	{%
		\includegraphics[scale=.32]{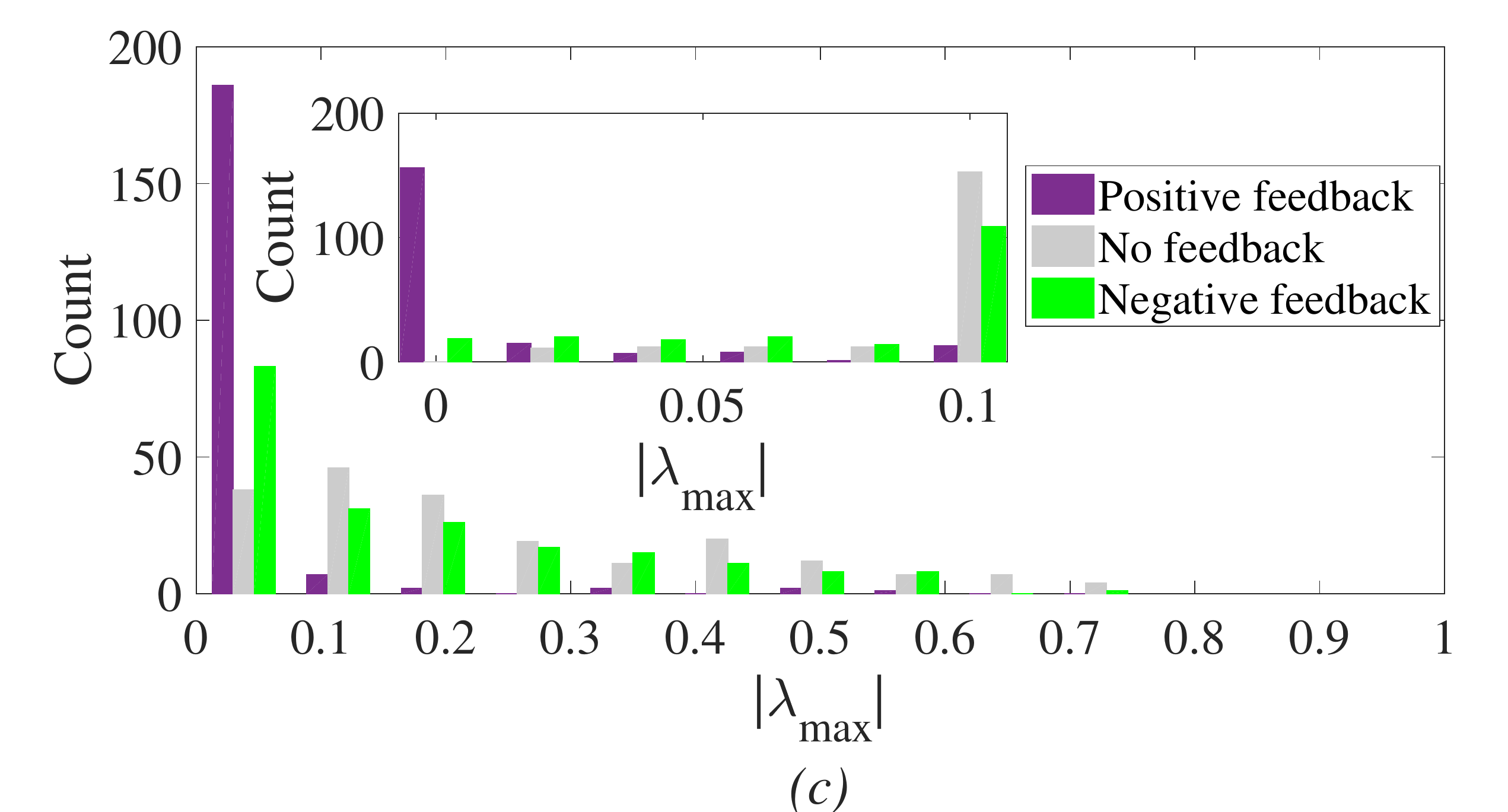}
		\label{strength50}
		
		}
		~

	\caption{Structural modifications in repressilator can improve its robustness to pulse perturbation.
	\fontsize{7}{7}\selectfont{\textit{(a)} Schematic showing two possible ways to modify the repressilator structure. \textit{(b)} Bar plot showing the distribution of dominant Floquet multipliers obtained for 100 periodic orbits of 3 node ($N=3$) and 5 node ($N=5$) circuits. The inset shows the distribution of the dominant Floquet multiplier within the range \numrange[range-phrase=--]{0}{0.1}. The parameter sets in both cases are chosen from a nominal parameter range of the repressilator. \textit{(c)} Bar plot showing the distribution of dominant Floquet multipliers obtained for 200 periodic orbits each for three different feedback models. The distribution with violet bars is for the positive feedback model, the grey bar for the no feedback model, and green bar for the negative feedback model. The inset within the figure shows the distribution of dominant Floquet multiplier within the range \numrange[range-phrase=--]{0}{0.1}. The range of parameters chosen for the feedback models are: $\alpha$ =\,\numrange[range-phrase = --]{2}{20}, $\alpha_{s}$ =\,\numrange[range-phrase = --]{5}{50}, $\alpha_{n}$ =\,\numrange[range-phrase = --]{5}{50}, $K$ =\,\numrange[range-phrase = --]{1}{10}, $J$ =\,\numrange[range-phrase = --]{100}{500}, $n$ = 2, $m$ = 2,  $\delta$\,=\,\numrange[range-phrase = --]{.2}{.5}, $k$\,=\,\numrange[range-phrase = --]{1}{10}, $\gamma$\,=\,\numrange[range-phrase = --]{0.01}{.1}}
}
\end{figure*}
\subsubsection{Adding positive feedback}

Another way to increase delay is to add positive feedback to an oscillatory circuit~\cite{novak2008design,ananthasubramaniam2014positive}. In fact, an approximate technique was used to quantitatively estimate the amount of delay provided by the positive feedback~\cite{dey2017describing}. To check such robustness for the present reressilator model, we compared three circuits: repressilator  with additional positive feedback repressilator with no additional feedback and repressilator with additional negative feedback (Figure 4\textit{(a)}). These modifications were modeled by changing the transcriptional function of $m_{P_{1}}$ in (1) to the following
\begin{equation} 
\begin{aligned}
\dot{m}_{P_{1}}=F(P_{1},P_{3})-\delta\,m_{P_{1}},
\end{aligned}
\end{equation}
Here, the additional positive and negative feedback is modeled through a modified Hill function $F$. For the additional positive feedback model, $F(P_{1},P_{3})=\alpha_{0}+\dfrac{\alpha_{s}\,\left(P_{1}/J\right)^{m}+\alpha}{1+\left(P_{3}/K\right)^{n}+\left(P_{1}/J\right)^{m}} $. The parameters $\alpha_{s}$ and $\alpha$ are the promoter strengths, $K$ and $J$ are the dissociation constants and $m$ and $n$ are the Hill coefficients. For the additional negative feedback model, $F(P_{1},P_{3})=\alpha_{0}+\dfrac{\alpha_{n}}{1+\left(P_{3}/K\right)^{n}+\left(P_{1}/J\right)^{m}}$, where $\alpha_{n}$ is the promoter strength and the parameter $K$, $n$, $J$ and $m$ denote the same interaction as above. For the no additional feedback model, the original model equations (1) is considered.

Using a similar approach as mentioned in Sec \ref{ring length}, we compare these models by exploring their parameter spaces randomly and compare their distribution of the Floquet multipliers.  In this case we select 200 common  parameter sets which generates oscillations in all the  three circuits. On plotting the histogram plot (Figure 4\textit{(c)}) of the Floquet multipliers we find a large fraction of the oscillations in case of the positive feedback circuit have a smaller Floquet multiplier value compared to the others. On further analysing the frequency distribution between the range \numrange[range-phrase = --]{0}{0.1} (shown in the inset of Figure 4\textit{(c)}), we find that most of the oscillations have Floquet multipliers which are an order of magnitude smaller than that of no feedback and negative feedback oscillations . This provides strong evidence that adding  positive feedback improves robustness to pulse perturbation.

We note that an increase in the number of stages in this manner as well as adding positive feedback has been found to enhance robustness to various perturbations due to variation in the parameters and temperature~\cite{stricker2008fast}.

\vspace{-1mm}
\section{Discussion}\label{sec14}
Robustness to transient perturbations may be an important requirement for a biomolecular oscillator. Using computations based on Floquet multipliers, we investigated the settling time after a pulse perturbation in the oscillatory dynamics of the repressilator, a benchmark biomolecular oscillator, and present the three following results. First, we found a strong correlation between the settling time estimated by the dominant Floquet multiplier and the settling time computed directly from nonlinear simulation, both at small and large pulse amplitudes. Second, we investigated the dependence of the Floquet multipliers on the circuit parameters, finding evidence that those variations that increase the effective delay, such as comparable mRNA and protein timescale or larger Hill coefficients, can reduce the settling time. Third, we considered structural modifications to the repressilator such as increasing the number of stages in the oscillator ring and additional positive feedback, and found that these can also reduce the settling time after pulse perturbation. These results highlight the important role of effective delay in oscillator dynamics in enhancing robustness to pulse perturbations.

It is interesting to note the usefulness of the dominant Floquet multiplier in assessing robustness of oscillator to pulse perturbation. First, it appears to capture the effective delay in the oscillator, at least from the point of view of degradation constants of the mRNA and protein as well as the Hill coefficient. Such a delay may buffer the oscillator from transient perturbations. Second, the settling time estimated from the Floquet multiplier correlates well with the settling time computed directly, both at small and large amplitude perturbations providing a convenient metric.
Third, the typical computational time required to obtain the Floquet multiplier can be smaller than the time to compute the settling time directly. Fourth, the Floquet multiplier can be inherently normalized to provide an estimate of the settling time relative to the period. For example, consider two oscillator with same settling time, but different periods. The oscillator with larger period is more robust as its settling time is a smaller proportion of its time period. The Floquet multiplier metric automatically captures this feature. Fifth, the Floquet method provides an analytical framework for the computational investigation of robustness to pulse perturbations.

A potential direction of future work is to see whether multiple feedbacks (double-positive or double-negative) or more complex motifs can similarly improve robustness to pulse perturbation. Further, oscillator with feedbacks implemented through other protein-based mechanism such as sequestration~\cite{venturelli2012synergistic} and degradation~\cite{xu2012roles} may be investigated from the point of view of robustness to pulse perturbation. Finally, the present analysis uses deterministic equations that are valid in the limit of small intrinsic noise. A consideration of how intrinsic noise, impacts robustness in such contexts may help present a complete picture, especially as such noise may change the dynamics.

In conclusion, the result presented here provide guidelines for the design of oscillators that are robust to pulse perturbations. In particular, they highlight the role of effective delay in an oscillator in facilitating such robustness. These results also provide an enhanced understanding of the transient response of an oscillator subjected to a dynamic input through the use of Floquet multiplier of an oscillatory trajectory. This may be relevant to control of oscillator dynamics where it is desired to switch oscillator features like time period from one value to the other.
\section{Methods}\label{sec14}

\subsection{Floquet theory}\label{floq_theory}
Floquet theory is an important method to probe the stability of limit cycle oscillations~\cite{strogatz2018nonlinear}. To illustrate it consider a general system governed by the dynamics,
\begin{equation}\label {non_linear_eq} 
 \dot{\boldsymbol{\mathbf{x}}}=f(\boldsymbol{\mathbf{x},\sigma}),
\end{equation}
where $\mathbf{x}=[x_{1}\;\;x_{2}\;\;x_{3}.......x_{n}]^{T}$ are the state variables representing the physical quantities of the system evolving in time and  $\boldsymbol{\mathbf{\sigma}}$ represents the parameter set. Suppose for a certain parameter set, there is a limit cycle $\mathbf{x}(t)=\mathbf{x_{P}}(t)$ with an oscillation period $T$. The stability and other properties of this limit cycle can be obtained by linearising the nonlinear equation (5) around the limit cycle $\mathbf{x_{P}}(t)$ as,
\begin{equation}\label {linear_eq} 
\Delta\dot{\mathbf{x}}=\mathbf{A}\Delta\mathbf{\mathbf{x}},
\end{equation}
where $\Delta\mathbf{x}$ is the perturbation and the Jacobian is $\mathbf{A}=\frac{\partial\mathbf{f}}{\partial\mathbf{x}}\biggr\rvert_{\mathbf{x}=\mathbf{x_{P}}(t)}$.

As a standard result of the Floquet theory~\cite{klausmeier2008floquet}, the perturbation $\Delta\mathbf{x}$ can be mathematically expressed as

\begin{equation}
\Delta\mathbf{x}(t)=\sum_{i=1}^{n}c_{i}\;e^{\mu_{i}t}\;\mathbf{p}_{i}(t), 
\end{equation}

where $\mathbf{p}_{i}(t)$ are the time periodic functions of period $T$  and $c_{i}$ are the constant coefficient and $\mu_{i}$ are the characteristic or the Floquet exponents. These exponents govern the decay (or growth) rate of the perturbations. A positive exponent means that the perturbation grows with time and a negative exponent means that the perturbation decays. These rates can also be analysed using the characteristic or Floquet multipliers which relates to the Floquet exponents as $\lambda_{i}=e^{\mu_{i}T}$. For a stable limit cycle, the modulus value of these Floquet multipliers lie between 0 and 1, except for one multiplier which is always unity. Physically, these multipliers signifies how fast a perturbation will decay over a complete period. Smaller multiplier means faster decay. To illustrate this, for a stable limit cycle, we first divide the perturbation $\Delta\mathbf{x}$ into two parts - one tangential to the limit cycle ($\Delta\mathbf{x}_{p}$) and the other  perpendicular to the cycle ($\Delta\mathbf{x}_{n}$),

\begin{equation}
\Delta\mathbf{x}(t)=\Delta\mathbf{x}_{p}(t)+\Delta\mathbf{x}_{n}(t)
\end{equation}
Here, $\Delta\mathbf{x}_{p}(t)=c_{p}\mathbf{\,p}_{p}(t)$ and $\Delta\mathbf{x}_{n}(t)=\Delta\mathbf{x}(t)-\Delta\mathbf{x}_{p}(t)$, where $c_{p}$ and $p_{p}$ are the coefficient and periodic function respectively corresponding to the tangential part of the perturbation. Let $\mu_{min}$ be the smallest Floquet exponent which dominate the relaxation of $\Delta\mathbf{x}_{n}(t)$. Assuming that the other terms decay faster than the terms containing $e^{\mu_{min}t}$ , $\Delta\mathbf{x}_{n}(t)$ can be approximated as 
\begin{equation}
\Delta\mathbf{x}_{n}(t)\cong c_{min}\,e^{\mu_{min}t}\;\mathbf{p}_{min}(t)
\end{equation}
where $c_{min}$ is the constant coefficient and $\mathbf{p}_{min}$ is the periodic function corresponding to $\mu_{min}$. The above perturbation obtained after one cycle will then be
\begin{equation}
\Delta\mathbf{x}_{n}(t+T)\cong c_{min}\,e^{\mu_{min}\left(t+T\right)}\,\mathbf{p}_{min}(t+T)=e^{\mu_{min}T}\,\Delta\mathbf{x}_{n}(t)
\end{equation}

\noindent where $\lambda_{max}$ is the dominant Floquet multiplier which approximately quantifies how fast a perturbation away from the limit cycle will decay over a complete period. From this, we can estimate the time the perturbation will take to settle using the following relation $\tau=\frac{1}{\mu_{min}}=\frac{T}{ln(1/|\lambda_{max}|)}$.

\subsection{Floquet multiplier computation}\label{floq_theory}
The Floquet multipliers are usually computed numerically using the following standard procedure~\cite{klausmeier2008floquet}. We start by solving (5) numerically to obtain the limit cycle and its corresponding period $T$. Next, we consider the  matrix differential equation given by
\begin{equation}\label {linear_eq} 
\dot{\mathbf{X}}=\mathbf{A}\mathbf{\mathbf{X}},
\end{equation}
where $\mathbf{X}$ is the fundamental matrix containing the basis vectors of which the perturbation $\Delta\mathbf{x}$ is composed of and $\mathbf{A}$ is the Jacobian matrix used earlier in (6). The above equation is solved for one time period starting from an initial condition $\mathbf{X}(0)=\mathbf{I}$. On solving we obtain the fundamental matrix $\mathbf{X}(T)$ whose eigenvalues are the required Floquet multipliers.  

The dominant Floquet multiplier computed here have its values accurate up to an error bound which is $10^{-3}$. We have computed this bound from the standard deviation of the unity multiplier values obtained from different parametric conditions. Due to this limitation, we have not presented the Floquet multiplier data for the ring oscillator model for $N$ larger than five. As the values are below the error margin, the comparison is not valid.

\section{Acknowledgements}
 We thank  Mr Abhilash Patel, Dr Venkat Bokka and Dr Abhishek Dey for their valuable suggestions. S.B. would like to thank the Visvesvaraya PhD Scheme for Electronics \& IT, Department of Electronics and Information Technology (DeitY), Government of India, for financial support (File No. IITD/IRD/MI01233).

%
%
%


\end{document}